\documentclass[letter,conference]{IEEEtran}
\IEEEoverridecommandlockouts
\usepackage{xspace}
\usepackage{booktabs}
\usepackage[most]{tcolorbox}
\usepackage{bbding}
\usepackage{graphicx}
\usepackage{subfig}
\usepackage{float}
\usepackage{multirow}
\usepackage{comment}
\usepackage{url}
\usepackage{hyperref}
\usepackage[]{mdframed}

\newtcolorbox{cooltextbox}[1][]{%
    colback=black!5,
    colframe=black!5,
    notitle,
    sharp corners,  
    borderline west={0pt}{0pt}{blue!80!black},
    enhanced,
    breakable,
    left=0pt,
    right=0pt,
    top=0pt,
    bottom=0pt
}

\newtcolorbox{cooltextbox2}[1][]{%
    colback=black!5,
    colframe=black!5,
    notitle,    
    enhanced,
    breakable,
    left=0pt,
    right=0pt,
    top=0pt,
    bottom=0pt
}

\title{Multimodal Large Language Models for Phishing Webpage Detection and Identification}

\begin{document}
\author{
    \IEEEauthorblockN{Jehyun Lee\IEEEauthorrefmark{1},
                    Peiyuan Lim\IEEEauthorrefmark{2},
                    Bryan Hooi\IEEEauthorrefmark{2},
                    Dinil Mon Divakaran\IEEEauthorrefmark{3}}
    \IEEEauthorblockA{\IEEEauthorrefmark{1}Trustwave,
        \IEEEauthorrefmark{2}National University of Singapore,
        \IEEEauthorrefmark{3}A*STAR Institute for Infocomm Research\\
        Email: jehyun.lee@trustwave.com,
        peiyuanlim@u.nus.edu,
        bhooi@comp.nus.edu.sg,
        dinil\_divakaran@i2r.a-star.edu.sg
    }
}

\maketitle

\begin{abstract}

To address the challenging problem of detecting phishing webpages, researchers have developed numerous solutions, in particular those based on machine learning (ML) algorithms. Among these, brand-based phishing detection that uses models from Computer Vision to detect if a given webpage is imitating a well-known brand has received widespread attention. However, such models are costly and difficult to maintain, as they need to be retrained with labeled dataset that has to be regularly and continuously collected. Besides, they also need to maintain a good reference list of well-known websites and related meta-data for effective performance. 

In this work, we take steps to study the efficacy of large language models (LLMs), in particular the multimodal LLMs, in detecting phishing webpages. Given that the LLMs are pretrained on a large corpus of data, we aim to make use of their understanding of different aspects of a webpage (logo, theme, favicon, etc.) to identify the brand of a given webpage and compare the identified brand with the domain name in the URL to detect a phishing attack. We propose a two-phase system employing LLMs in both phases: the first phase focuses on brand identification, while the second verifies the domain. 
We carry out comprehensive evaluations on a newly collected dataset. Our experiments show that the LLM-based system achieves a high detection rate at high precision; importantly, it also provides interpretable evidence for the decisions. Our system also performs significantly better than a state-of-the-art brand-based phishing detection system while demonstrating robustness against two known adversarial attacks. 

\end{abstract}

\section{Introduction}
Phishing attacks have become increasingly prevalent in our digital world, posing a major threat to individuals, businesses, and governments. According to a recent report from Anti-Phishing Working Group (APWG), nearly 1~million unique phishing webpages were observed for the first quarter of 2024~\cite{apwg2024report}.
In response to this evolving cyber threat, researchers have made significant efforts to develop effective techniques for detecting phishing webpages. In particular, to overcome the limitations of blacklists in detecting new phishing webpages, numerous solutions utilizing machine learning (ML) approaches have been proposed~\cite{divakaran2022phishing}.

Conventional ML-based approaches train a supervised model to learn and differentiate the characteristics of the two classes of webpages, namely \texttt{benign} and \texttt{phishing}, e.g., ~\cite{Google-content-phishing-2010,canita+-2011, li2019stacking, lee2020building}. 
However, such approaches face a critical challenge as phishing pages become more sophisticated and reuse resources from benign pages they target~\cite{lim2024phishing}. This attacker strategy increases the similarity between phishing and benign pages, making detection increasingly difficult. The similarity of a phishing page to its target legitimate page serves the primary goal of an attacker, as that helps to deceive the victims. It is in this context that the \textit{brand-based} phishing detection techniques are relevant~\cite{abdelnabi2020visualphishnet, lin2021phishpedia, liu2022inferring, hout2022logomotive,li2024knowphish,ji2024evaluating}. 

Brand-based phishing detection approaches are developed relying on the empirical evidence that most phishing webpages attempt to imitate popular and trusted brands. As per statistics~\cite{lin2021phishpedia, liu2022inferring, apwg2024report}, more than 90\% phishing attacks target just a couple of hundreds of brands. Therefore, brand-based phishing detection solutions try to identify the brand that a given webpage tries to imitate or render. Subsequently, if the domain name in the URL is different from the identified brand, then the page is likely to be a phishing page. Brand-based phishing detectors have dual advantages---besides detecting a phishing page, they are also able to identify the targeted brand. Recent brand-based phishing detectors use well-known models from Computer Vision (CV), analyzing a webpage  screenshot~\cite{fu2006detecting, abdelnabi2020visualphishnet} or the logos on screenshots~\cite{lin2021phishpedia, liu2022inferring}, to identify the brand a webpage is rendering.

The models in the most recent brand-based phishing detection proposals, in particular VisualPhishNet~\cite{abdelnabi2020visualphishnet} and Phishpedia~\cite{lin2021phishpedia}, are trained to protect a reference list of well-known brands. This requires labeled dataset of phishing and benign webpages. In fact, Phishpedia also requires the coordinates of the logos on webpages to be indicated in the dataset to train its object (i.e., logo) detection model. Note that datasets must be collected not just once but continuously at regular intervals to retrain the models so that the detection systems adapt to the changing dynamics of phishing attacks and their target (benign) pages over time. Clearly, this process 
is laborious and costly. Furthermore, maintaining a reference list for effective domain verification with the identified brand is also challenging~\cite{dynaPhish-2023, li2024knowphish}.

Recent times have witnessed the rise of large language models (LLMs) pretrained on massive datasets and capable of performing multiple downstream tasks. The underlying transformer architecture~\cite{vaswani2017attention} of the LLMs has enabled the creation of powerful models with billions and even trillions of parameters. 
Among these, we believe the multimodal LLMs, which can analyze various data types such as text and images, show significant potential for brand-based phishing detection.  With their vast knowledge, the multimodal LLMs would have learned about webpage representations of numerous well-known brands.  Therefore, these LLMs may have the capability to identify brands of a given webpage, by analyzing the logos present, the theme of the page, the contents, etc.

In this study, we take a fresh look at phishing detection and explore the potential of multimodal LLMs in {\it automatically} detecting phishing pages through brand recognition. 
The biggest advantage of having an LLM-based phishing detection system is that it does not require any labeled data.
This also allows us to build a solution that is no more limited to one type of data, e.g., logos~\cite{lin2021phishpedia}, or HTML contents~\cite{canita+-2011}, or screenshots~\cite{abdelnabi2020visualphishnet}. 
Furthermore, LLMs offer the opportunity to extract and provide interpretations of results, a capability lacking in existing research proposals. This interpretability could significantly enhance the understanding and effectiveness of phishing detection mechanisms.

Specifically, we try to answer the following questions:

\begin{itemize}
    \item How do LLMs perform in detecting phishing webpages and in identifying the targets of the phishing attacks, and how good are they in providing explanations? ({Sections~\ref{sec:eval:llm}},~\ref{subsec:second-phase-LLM} and~\ref{subsec:interpretation})

    \item Besides the screenshot, HTML contents are also available from a webpage. What are the gains, synergy, and conflicts due to the different inputs (screenshot/HTML) and their combination in phishing detection performance? ({Sections~\ref{sec:eval:llm}},~\ref{subsec:exclusive-LLM} and~\ref{sec:eval:conflict}) 

    \item How does the LLM-based phishing detection system perform in comparison to a state-of-the-art reference-based phishing detection solution? ({Section~\ref{sec:eval:vpn}})

    \item Using the APIs of pretrained LLMs for inference tasks comes with a cost. What is the cost incurred in using an LLM-based phishing detection system? ({Section~\ref{sec:eval:cost}})

    \item Recent works have shown that visual-based phishing systems can be evaded using adversarial perturbations~\cite{apruzzese2023position,lee2023attacking,kulkarni2024-mlllm-evaluating-robustness}. How robust are LLM-based phishing systems in response to these adversarial attacks? ({Section~\ref{sec:eval:adv}})

    \item What are the new attack vectors due to the use of the large pretrained models for phishing detection? ({Section~\ref{sec:discussion}})

\end{itemize}

To answer the above questions, we first carefully collect a new phishing dataset with manually labeled attack targets. We design and develop a new system that processes given webpages and utilizes multimodal LLMs to, both, detect and identify phishing webpages ({Section~\ref{sec:system}}). We evaluate three state-of-the-art multimodal LLMs, namely GPT-4, GeminiPro~1.0, and Claude3, on their capability to assist with phishing detection. Our experiments demonstrate that LLMs show promise in detecting phishing pages at high precision, while also providing explanations. Equally importantly, we study the robustness of our system against adversarial phishing samples and observe that the LLM-based design helps in successfully identifying the target brands even in the face of adversarial perturbations. 

We make the following contributions through this work:

\begin{itemize}
    \item We build an end-to-end pipeline that uses LLMs for phishing detection, identification, and result explanation. During the process, we also define prompts that are effective in achieving high detection accuracy~({Section~\ref{sec:system}}). We publish the code-base online~\cite{LLM-src-code}.
    
    \item We collect, process, and analyze a new dataset to evaluate phishing detection solutions ({Section~\ref{sec:data}}). We publish both the dataset and the system for automatically collecting new datasets, so as to make it easier for researchers to gather datasets in future~\cite{LLM-src-code}.
    
    \item We carry out comprehensive performance evaluations of the LLM-based phishing detection system using the latest dataset we collected. We evaluate the system using three LLMs, across different input types---HTML, screenshot, and both, and present the results ({Section~\ref{sec:eval}}). We also present a case study on webpages that are wrongly identified (Section~\ref{sec:case-study}).
    
    \item As LLMs offer new capabilities for phishing detection, we also study corresponding relevant aspects:

    \begin{itemize}
        \item We evaluate the cost of using LLMs for the purpose of phishing detection ({Section~\ref{sec:eval:cost}}).

        \item We evaluate the robustness of LLM-based phishing detection to recently proposed adversarial attacks against phishing models ({Section~\ref{sec:eval:adv}}).
        
        \item We discuss new attack vectors arising from the use of LLMs ({Section~\ref{sec:discussion}}).
    \end{itemize}
\end{itemize}
\section{Related works}

Without being exhaustive, we review some of the relevant existing research works below. 

\subsection{HTML-based approaches} Earlier works have explored the possibility of using HTML in phishing detection.  HTML-based approaches have focused on HTML tags for malicious functionalities and having a similar layout to the target webpages and structural characteristics of phishing web pages.

Rosiello~\textit{et al.} proposed DOMAntiPhish~\cite{rosiello2007layout} that compares the DOM structure of a webpage to the previously stored webpages where a user inputs sensitive information.  If the same information is input to a webpage with a similar layout in terms of DOM structure, it is assumed to be a phishing attempt.  This study is based on the rationale that a phishing webpage tries to be similar to the target benign webpage. However, it faces limitations against the phishing pages which are only visually similar to the genuine web page but different in DOM structure.  Li~\textit{et al.}  proposed a stacking model leveraging URL and HTML features~\cite{li2019stacking}. The model introduced HTML string embedding and crafted lightweight features from URLs and HTML scripts to detect phishing pages.
Multiple studies have shown the vulnerability of HTML-based solutions~\cite{lee2020building, apruzzese2022spacephish}.
Lee~\textit{et al.} showed that tree-based models are prone to evasion attacks, where attackers can easily manipulate the most relevant features in the model~\cite{lee2020building}. Subsequently, they proposed a phishing detection system that reduces the reliance on a few specific features by adding noise to the data. Similarly, Chiew~\textit{et al.} proposed a hybrid ensemble feature selection framework for an ML-based phishing detection system~\cite{chiew2019new}. Saleh~\textit{et al.}~\cite{saleh2021identi} proposed a method for identifying and detecting phishing attacks by using deep learning models on the textual content of websites.

\subsection{Visual-based approaches}
Visual-based phishing webpage detection uses the visual features or invariants of a webpage to identify and classify it as a phishing website~\cite{fu2006detecting, phishzoo-2011, liu2017phishing, logosense-2020, abdelnabi2020visualphishnet, lin2021phishpedia, 
hout2022logomotive, liu2022inferring}. Such techniques rely on analyzing the layout, design, and images used on a webpage to differentiate between legitimate and malicious websites.
In one of the earliest works, Fu~\textit{et al.}~\cite{fu2006detecting} proposed a phishing web page detection system by measuring Earth Mover's Distance (EMD) between screenshots of webpages.  Their approach directly uses colors and coordinates of an image for building an image signature in contrast to the subsequent studies that used neural networks as the deep learning model made significant advancements. 

Liu~\textit{et al.}~\cite{liu2017phishing} proposed a Convolutional Neural Network (CNN) based approach for detecting phishing webpages. They extracted visual features from the captured screenshots of webpages and used a CNN classifier to distinguish between legitimate and phishing pages.  VisualPhishNet, proposed by Abdelnabi~\textit{et al}.~\cite{abdelnabi2020visualphishnet} uses an advanced approach by training a CNN with triplet samples that utilize the multiple webpage screenshots from the same and different brands, with the goal of identifying a webpage that looks similar to a protected website (a small list of legitimate websites). 

Logo-based phishing detection is a type of visual-based technique that specifically focuses on detecting and identifying the logo of a website as part of the phishing detection process. The advancement of models in Computer Vision has helped develop practical solutions in this space. Lin~\textit{et al.} proposed Phishpedia~\cite{lin2021phishpedia},  a system that combines Faster-RCNN and Siamese models to detect and identify logos on webpage screenshots. Narwaria~\textit{et al.} proposed a framework for detecting phishing URLs, PhishAOD~\cite{narwaria2020phishaod}, by analyzing the visual similarity between the logo of the given webpage and that of legitimate ones. They used a deep learning approach to extract features from the logos.
Tan \textit{et al.} proposed a hybrid approach using visual and textual identities of a webpage to overcome the performance limitation of previous visual-only approaches~\cite{tan2023hybrid}.

\subsection{LLMs for security}
LLMs are currently being explored for solving challenges in the domain of security as well as online trust and safety~\cite{divakaran-LLM-security-2024}. AutoCodeRover~\cite{zhang2024autocoderover} uses multiple LLMs to locate bugs in codes and generate patches for GitHub issues. 
He~\textit{et al.}~\cite{he2023you} propose using LLMs with prompt-learning capabilities to detect toxic webpage contents. 
LLMs have also been used for phishing email detection~\cite{d-fence-2021, koide2024chatspamdetector}.
Finally, in a recent proposal, LLM is utilized to build knowledge-base of brands for brand-based phishing detection solutions~\cite{li2024knowphish}.
\section{LLM-based Phishing webpage detection}\label{sec:system}
In this section, we describe the design of our phishing webpage detection system utilizing LLMs.

\subsection{System design}
\begin{figure}[tbp]
    \centering    
    \includegraphics[width=0.47\textwidth]{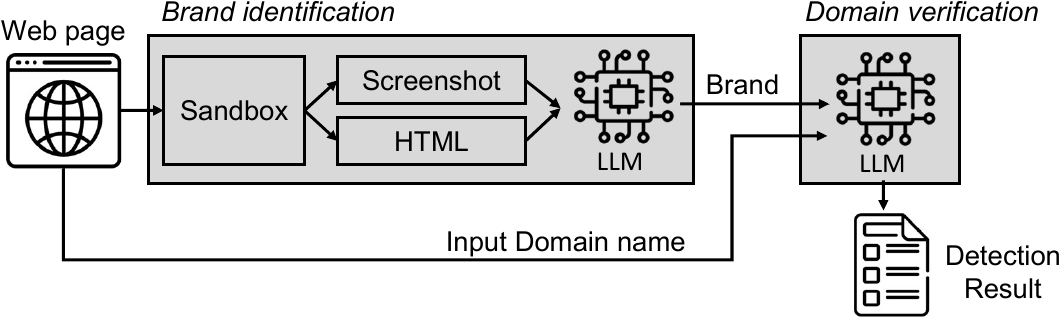}
    \caption{System overview of LLM-based Phishing webpage detection system }\label{fig:overview}
\end{figure}

A brand-based detection system works in two phases:
\begin{itemize}
    \item Brand identification: Identify the brand of a given webpage, based on its logos, theme, description, etc.
    
    \item Domain verification: Check whether the domain name of the given page {\em matches} the identified brand.
\end{itemize}

Much of the modeling efforts in the recent brand-based phishing detection proposals~\cite{abdelnabi2020visualphishnet, lin2021phishpedia, liu2022inferring} have been to solve the first challenge of brand identification. The Computer Vision models used in these proposals look for visual similarities with existing well-known brands. Once the brand is identified, the second and final task is cross-checking the URL and matching the domain name with the identified brand. For example, if the brand identified is that of Paypal, the domain name in the URL is expected to be \texttt{paypal.com}, if the webpage is legitimate. However, such a simple comparison comes with pitfalls:
\begin{itemize}
    
    \item Brand names might change over time (Facebook to Meta).   
    
    \item Brand name and domain name might look legitimately different, or a brand might have aliases or sub-brands (x.com and twitter.com). 
    
    \item Brands may be represented differently in various regions or languages. (e.g., Zuimeitianqi and Zuimei weather)
    
    \item Often, webpages carry product information (e.g., iPhone, Outlook, etc.) that might be extracted as the best semantic feature by a machine learning model rather than the brand itself.
    Therefore, there is also a subtle challenge of mapping products to brands.
    
    \item Models might also extract a trademark instead of a brand name. 
    
    \item As a final point, the list of brands targeted by phishing changes over time~\cite{dynaPhish-2023, li2024knowphish}. 

\end{itemize}

Given these challenges, maintaining a static list of domain names corresponding to a set of reference brands is not an effective solution.
Considering the above points, we design our system in two phases of {\em brand identification} and {\em domain verification} and utilize LLMs for both phases. 
{Figure~\ref{fig:overview}} illustrates the system.

\paragraph{\textbf{Brand identification}}  First, a given webpage is introduced into a sandbox environment to ensure safety from potentially malicious content, wherein both a screenshot and HTML content of the page are extracted from the rendered webpage via a web engine. 
The HTML contents are processed to extract relevant information such as the title, meta description, path of the favicon, text corresponding to the logo images, header and footer texts, etc. Note, we do this preprocessing primarily to reduce the input token size and keep the cost manageable (refer Section~\ref{sec:eval:cost}).
These two outputs, the visual representation via the screenshot and the textual data extracted from the HTML,  are subsequently provided to a multimodal LLM. Using an engineered prompt, the LLM is instructed to process these inputs to identify the brand (detailed below in {Section~\ref{sec:system:prompt}}). We also instruct the LLM to provide output in a particular format. The output includes the identified brand, the brand-related elements on the page, and, equally importantly, the supporting evidence. The following example illustrates the output from an LLM. 

\paragraph{\textbf{Phishing webpage classification}}
Once the brand is recognized, the identified brand information is passed through to another LLM. This additional layer of LLM is instructed to compare the identified brand and the domain name in the given URL to carry out domain verification. As LLMs have learned from massive amounts of online data, we believe they are equipped to understand the association between brand name and domain name, overcoming the challenges mentioned above.  
The final output is a detailed phishing detection result, which encompasses comprehensive insights regarding the identified brand per the input domain name. We demonstrate this in Section~\ref{subsec:interpretation}.

\subsection{Prompts for brand identification}\label{sec:system:prompt}

We design the prompts to provide instructions along with screenshots and/or HTML  information to LLMs, so as to determine the target brand of the given webpage.  Besides pinpointing the webpage's target brand, we further extend the task to include the identification of any visible fields soliciting credentials, e.g., passwords or identification numbers, call-to-action buttons, e.g., \textit{Login} or \textit{Click here to proceed}, as well as supporting evidence.  This comprehensive approach aims to harness LLM's analytical capabilities for subsequent evaluation, if necessary, and explainability of the results.

To examine three combinations of input data, we give different prompts to the LLMs depending on the data. Besides, there is a common instruction among all configurations. 
We first define the prompts for the three input data types: screenshot, HTML contents, and both.

\vspace{0.4cm}
\noindent \textit{{1. Prompt specifically with (only) webpage screenshot as input}}
\begin{cooltextbox2}
\small {
I want you to act as a webpage brand identifier.\\
Given a screenshot of a webpage. I want you to identify the brand of the webpage from the screenshot alone. \\
Additionally, also note whether fields are asking for sensitive user credentials as well as any call-to-action buttons / links.\\
Examples of sensitive user credentials: email, username, password, phone number, etc. \\
Examples of call-to-action elements: buttons or links that lead to asking for user credentials. \\

Return your response in the following format, and replace everything in [] with your answer:\\
1. \underline{Brand}: [response]\\
2. \underline{Has Credentials}: [Yes/No]\\
3. \underline{Has Call\_To\_Action}: [Yes/No]\\
4. \underline{List of credentials}: [response if Yes for (2), otherwise NA. Keep within the top 10 fields.]\\
5. \underline{List of call\_to\_action}: [response if Yes for (3), otherwise NA. Keep within the top 10 fields.]\\
6. \underline{Confidence Score}: [How confident are you when identifying the brand on a scale of 0.00 to 10.00 (in 2 decimal places), 10.00 being absolutely confident, 0.00 being not confident]\\
7. \underline{Supporting Evidence}: [response, keep it within 300 words]
}
\end{cooltextbox2}

\noindent \textit{{2. Prompt specifically with information from HTML as input}}
\begin{cooltextbox2}
\small{
I want you to act as a webpage brand identifier.\\
Given the key information from the HTML content of a webpage, I want you to identify the brand of the webpage from this information alone. The key information is extracted from the actual HTML script of the webpage, and this information includes the title of the webpage, the metadata, the favicon, the logo attribute, the footer and header text, and the nav bar. These areas are where the brand information can typically be found. Make good use of this information to identify the brand related to each webpage.\\

Return your response in the following format, and replace everything in [] with your answer::\\
1. \underline{Brand}: [response] \\
2. \underline{Confidence Score}: [How confident are you when identifying the brand on a scale of 0.00 to 10.00 (in 2 decimal places), 10.00 being absolutely confident, 0.00 being not confident]\\
3. \underline{Supporting Evidence}: [response, keep it within 300 words]
}
\end{cooltextbox2}

\noindent \textit{{3. Prompt with both screenshot and HTML as input}}
\begin{cooltextbox2}
\small {
I want you to act as a webpage brand identifier.\\
Given a screenshot and key information from the HTML script of a webpage. I want you to identify the brand of the webpage from these two sources alone.\\
Additionally, also note whether there are fields asking for sensitive user credentials as well as any call-to-action buttons/links.\\
Examples of sensitive user credentials: email, username, password, phone number, etc.\\
Examples of call-to-action elements: buttons or links that lead to asking for user credentials.\\

\noindent \textit{Response format the LLM is instructed to follow is the same as that for screenshot}
}
\end{cooltextbox2}

\noindent The following is the common prompt that is used along with all the above configurations to provide further instructions and restrict online access for safety purposes.
\begin{cooltextbox2} {\small
Please adhere strictly to the following rules for your analysis:\\
1. Do not interact with the webpage in a live environment or use browser functionalities.\\
2. Avoid inspecting the webpage's source code, the website's address bar, SSL certificates, URLs, or any interactive features.\\
3. Your analysis should be grounded solely on the given input data.\\
4. No additional resources or external validations should be used.\\

Note that some pages may include information / logos of other brands, especially pages that use single sign-on features or pages created by website builders.
}
\end{cooltextbox2}

\section{Data collection}\label{sec:data}
We collected our dataset by crawling the bengin and phishing webpages over a period of three months from Oct. to Dec. 2023. 
For benign pages, we visited the top-ranked websites from the Tranco list~\cite{pochat2019tranco}. Specifically, we crawled the top 10,000 ranked domains and those that ranked between 100,000 and 105,000, so as to include a broad spectrum of legitimate webpages that vary in popularity. Of these 15,000 pages, we applied filtering techniques (see {\bf Section~\ref{subsec:data:preprocessing}}) to remove invalid pages, domains such as googleapis.com, etc. Finally, we manually labeled the brands of these legitimate domain names limited by our budget. These processes brought down the labeled benign pages to around 3,000. 

For webpage crawling, we use Playwright~\cite{playwright2023} along with Python scripts; the preprocessing of collected data and invalid sample filtering are all implemented in Python.  
Specifically, we collected the screenshots using Playwright's screenshot API and extracted the essential information in HTML with the BeautifulSoup library of Python. We collected information such as title, favicon path/filename, logo image's alternate text, headers, footers, navigation bar content, paragraph texts, span text from the parsed HTML text, etc.

We used OpenPhish~\cite{openphish2023} for obtaining feeds of phishing URLs. 
Due to the volatility of phishing pages, authorities' mitigation efforts, and attackers' cloaking techniques, a significant number of phishing webpages collected from a crawler are inaccessible or invalid. On the other hand, adversaries often introduce multiple URLs that provide identical or similar web pages to maximize the success rate and viability of their phishing pages.
These duplicated and invalid samples in the dataset often lead to overestimation of detection performance by providing a model with many low-hanging fruits.

The following explains the process of obtaining labeled unique phishing webpages. 

\subsection{Dealing with anti-cloaking tactics of phishing pages}

\paragraph{{Preliminary Anti-cloaking test}} We performed a preliminary analysis to assess the cloaking behaviors of phishing pages based on previous studies~\cite{CrawlPhish, IntentionallyTriggeringCloakingBehavior}. This targeted experiment involving approximately 150 phishing URLs was conducted to determine the effects in phishing webpage rendering due to differences in environmental variations; specifically, the variations we tested include user-agent string in the browser, user interaction, referrer string, as well as multiple geolocations via VPN.

\begin{itemize}
  \item \textbf{User-agent}: According to our anti-cloaking testing with the environment variations in OSes (e.g., Windows and MacOS), device platforms (i.e., Desktop and Mobile devices), and crawling tools (i.e., Selenium and Playwright), it showed marginal variation between the different configurations, except for the explicit bot-like browser profiles.  Consequently, we adopt one of the most popularly used User-Agent strings that results in minimal cloaking, i.e., \texttt{Mozilla/5.0 (Windows NT 10.0; Win64; x64) AppleWebKit/537.36 (KHTML, like Gecko) Chrome/116.0.0.0 Safari/537.36}.
  
  \item \textbf{User interaction}: We tested different user actions upon opening the web pages, e.g., Page scrolling, Mouse clicking, and Mouse movement. We observed that only the integration of mouse movements showed a pronounced effect in evading cloaking techniques. Specifically, the configuration that had \ mouse movements recorded no blocking from the phishing pages we visited. 
  
  \item \textbf{Referrer}: We explored four referrers where users may encounter phishing URLs: Google, Facebook, phishing URL's own domain name, and an empty referrer. Our test results highlighted a notable increase in cloaking when no referrer was employed. In contrast, using a self-referential source led to a significant drop in the cloaking behavior.
\end{itemize}

\subsection{Preprocessing}
\label{subsec:data:preprocessing}
\paragraph{\textbf{Invalid sample filtering}} We ignore the empty and erroneous samples from the initial gathering based on:
\begin{itemize}
\item \textbf{Incomplete samples}: Samples with missing HTML scripts, screenshots, certificate info, DNS records, or network resources.
\item \textbf{Response Status}: Samples with non-200 HTTP response status codes.
\item \textbf{Semantic blank}: Samples with HTML contents and CSS styles that lead to potentially blank pages.
\item \textbf{Duplication}: Duplicated samples by applying SHA256 to URLs and HTML contents.
\item \textbf{Screenshot-pixel standard deviation}: Low standard deviation in grey-scale screenshot indicating a blank page.
\item \textbf{Screenshot-edge count}: Low canary edge count in grey-scale indicating fewer shapes and text, suggesting a lack of content.
\item \textbf{Screenshot-text length}: Short text detected by OCR indicating lack of significant contents.
\item \textbf{Verification or warning pages}: Screenshots that show security or human user verification, e.g., CloudFlare verification, or warnings pages by browser and ISP. Dealing with such techniques, e.g., CAPTCHAs, is complicated and outside the scope of this work~\cite{dai2024cframe}. 
\end{itemize}

A small-scale experiment was conducted to establish the threshold values for these metrics. This set includes examples of blank pages (in both white and black), nearly blank screenshots, screenshots with minimal content, and screenshots rich in content. 
The threshold values obtained from this experiment are used for the final dataset filtering.

\subsection{Brand labeling}
While OpenPhish feeds phishing webpages, we still need to identify and verify the brands targeted by the attack.
The verification process involves two stages. First, we used VirusTotal~\cite{virustotal}, a platform aggregating analyses from multiple security vendors, to evaluate URLs. We selected four vendors---OpenPhish, Google Safe Browsing, Kaspersky, and Trustwave. OpenPhish uses proprietary algorithms and machine learning algorithms; Google Safe Browsing benefits from extensive web crawling; Kaspersky is renowned for its cybersecurity research and development; and Trustwave provides solutions against phishing emails~\cite{d-fence-2021}. URL verification was conducted over three consecutive days to account for any changes or updates.

In the second stage, three analysts (including only one co-author) conducted manual reviews. They analyzed screenshots of each webpage and compared their respective brands to their URLs. This two-stage approach ensures high accuracy in categorizing samples as phishing or legitimate.
To classify a sample as phishing, both of these conditions had to be met: i) a webpage sample is flagged as phishing if all the four security vendors on VirusTotal over the three-day verification period. Previous research suggests that a URL flagged by at least two vendors typically confirms its phishing status, and raising this to four enhances robustness and minimizes false positives~\cite{lim2024phishing}, ii)~the sample was manually verified as phishing.

\subsection{Final Data }
Although we obtained around 377,000 phishing URL links from OpenPhish, based on the above filtering and labeling budget, our eventual set consists of 1,500 phishing pages with their target labels identified. Refer to the Table~\ref{tab:lmm_dataset} for the final numbers. 
However, we publish both the labeled and unlabeled data online~\cite{LLM-src-code} so that researchers can utilize them in the future. 

\section{Evaluation: Brand identification using LLM}~\label{sec:eval}

In this section, we present the evaluations of the LLM-based phishing detection system. We experiment with three commercial LLMs---{\textit{Google \underline{Gemini} Pro-Vision 1.0, OpenAI \underline{GPT}-4-turbo, and Anthropic \underline{Claude}3 Opus}}, as the first and second-phase LLM models in our detection system. 
Our goal here is to answer the following questions.\\

\begin{itemize}
    \item[\textbf{Q1.}] What is the performance of an LLM-based phishing system? 
    Which multimodal LLM shows better overall performance in phishing detection?

     \item [\textbf{Q2.}] Which input data (HTML/screenshot) helps in achieving higher accuracy?
    \item [\textbf{Q3.}] Does an LLM have any strengths/weaknesses with a specific input data type for the brand identification task? 
    
    \item[\textbf{Q4.}] What is the effectiveness of the second-phase LLM in detecting phishing webpages? 

    \item[\textbf{Q5.}] Does LLM help in providing interpretation of results? 
    
    \item [\textbf{Q6.}] What is the effect of one input data type over the other?

    \item [\textbf{Q7.}] How does LLM-based phishing detection perform when compared against a state-of-the-art solution? 
    
    \item [\textbf{Q8.}] What is the cost of using LLM for brand identification? How does it vary with the input types?
\end{itemize}

\paragraph*{\textbf{Dataset}} We use the manually labeled and verified benign and phishing webpage dataset (Table~\ref{tab:lmm_dataset}) described in {Section~\ref{sec:data}} for the evaluation of brand identification and phishing classification using LLMs.\\

\paragraph*{\textbf{Metrics}} We use the common metrics for evaluating the phishing detection capability of solutions. They include precision (that quantifies the misclassification), recall (or detection rate of phishing pages), and F1-score (harmonic mean of precision and recall). 

\begin{table}[t]
    \caption{LLM analysis dataset}
    \centering    
    \begin{tabular}{@{\hspace{1em}} l @{\hspace{1em}} | @{\hspace{1em}}  c @{\hspace{1em}} c @{\hspace{1em}}}    
    \toprule
      & Num. of Samples & Num. of Brands\\
    \midrule
    Benign         & 2981    & 1607  \\
    Phishing       & 1499    & 294   \\    
    \midrule
    Total          & 4480   & 1882   \\
    \bottomrule
    \end{tabular}
    \label{tab:lmm_dataset}    
\end{table}

\subsection{Phishing detection and target brand identification}\label{sec:eval:llm}

\paragraph{\textbf{Overall performance}} 

As mentioned earlier, our proposed system consists of two phases: brand identification and domain verification.

\begin{figure}
    \centering
    \subfloat[Precision]{\label{fig:prec_fig}
        \includegraphics[width=0.23\textwidth]{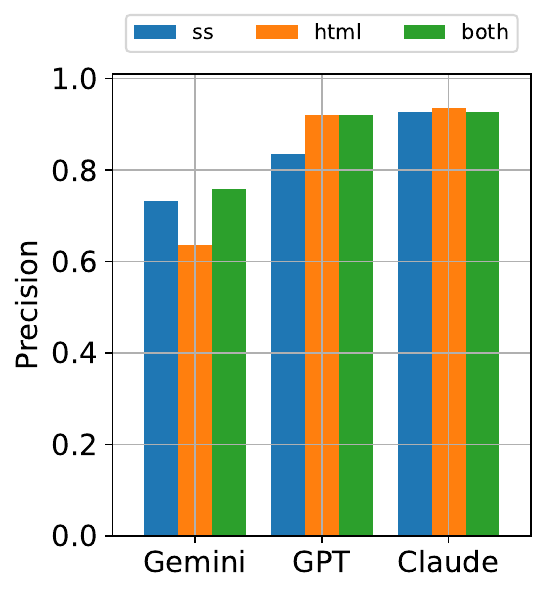}}
    \subfloat[Recall]{\label{fig:recall_fig}
        \includegraphics[width=0.23\textwidth]{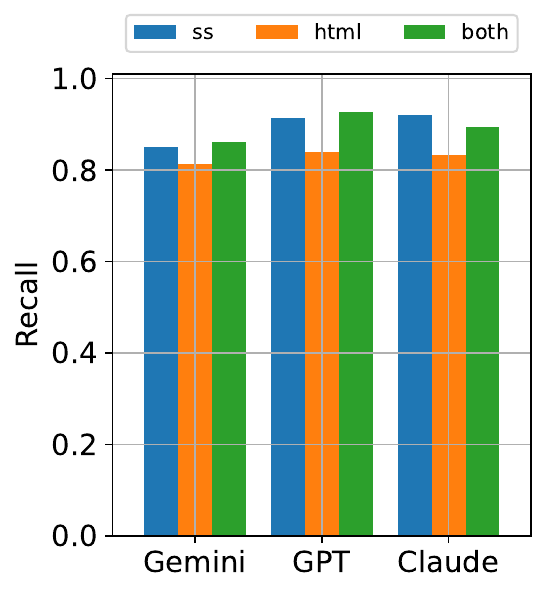}}
    \caption{Precision and Recall of phishing detection with LLMs}\label{fig:perform_llm_fig}
\end{figure}

{Figure~\ref{fig:perform_llm_fig}} presents the results for \textbf{Q1}, i.e., phishing detection performance involving both phases.

Let us analyze the case where  HTML contents and screenshots of webpages ({\bf both}) are given as input to the system. We observe that GPT and Claude3 have similar detection rates (recall) for almost the same precision values. The F1-score achieved by GPT-4 and Claude3 are approximately $0.92$ and $0.90$, respectively.  Gemini lags behind the other two---for Gemini to achieve a similar detection rate (as GPT and Claude), and the precision drops by more than 15\%. Consequently, the F1-score of Gemini is lower at $\approx0.81$.

{Figure~\ref{fig:perform_llm_fig}} answers \textbf{Q2} and \textbf{Q3} (also revisted later in \textbf{Section~\ref{subsec:exclusive-LLM}}). For Gemini and GPT, providing HTML content and screenshots together results in better precision and recall (and, therefore, the best F1-scores as well) than just either of them. Claude, however, is able to achieve the best performance with just a screenshot. Interestingly, the precision of Gemini drops to a low $63.6$\% when it is given only HTML as input. We discuss the reasons for misidentification and error cases in later sections below. 

\subsection{Effectiveness of domain verification (\textbf{Q4})}
\label{subsec:second-phase-LLM}

\begin{figure}
    \centering
    \subfloat[Precision]{\label{fig:prec_phase_fig}
        \includegraphics[width=0.23\textwidth]{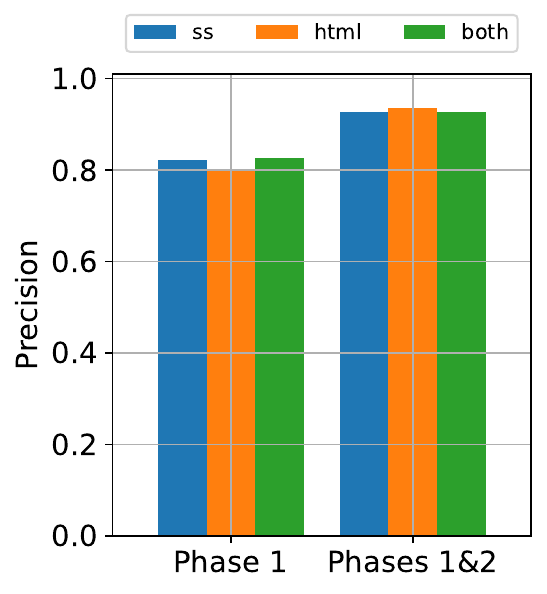}}
    \subfloat[Recall]{\label{fig:recall_phase_fig}
        \includegraphics[width=0.23\textwidth]{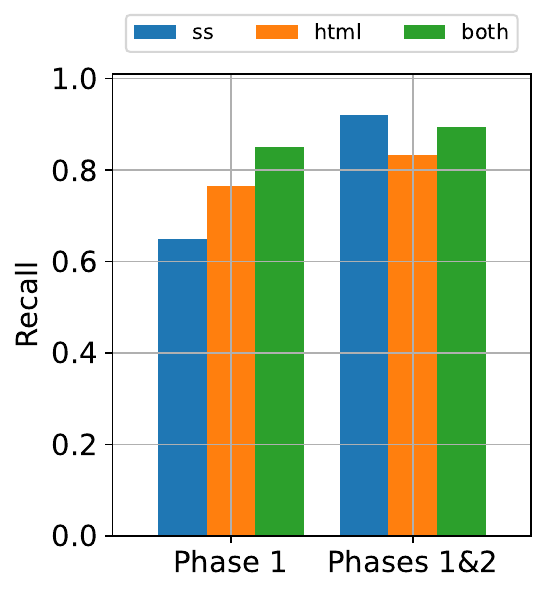}}
    \caption{Precision and Recall before and after the second-phase LLM. Results are for Claude.}\label{fig:perform_phase_compare}
\end{figure}

Next, we evaluate the effectiveness of the second phase of domain verification in the phishing detection system ({\bf Q4}). To do so, we remove the second-phase LLM and instead do a string comparison of the domain name to the brand identified by the first-phase LLM; this is denoted as `Phase~1' on Figure~\ref{fig:perform_phase_compare}. For the scenario where the whole system with both phases is in operation, we use the label `Phases~1\&2' on the same figure. The results presented are for Claude, each input type, and their combination. 

Since Claude achieves the highest detection rate (recall) with a webpage screenshot as input, let us consider the gain due to the second-phase LLM in this scenario. From Figure~\ref{fig:perform_phase_compare}, we observe that the second-phase LLM helped the precision to increase by more than 10\%, while the recall jumped from approximately 65\% to a high 92\%, resulting in an absolute increase of $17\%$ detection rate. This is reflected in the increase in F1-score, from $\approx73\%$ for Phase~1 to $\approx92\%$ when Phase~2 is added. Clearly, using the second-phase LLM to perform the comparison of the identified brand and domain name is better than doing a string matching. It is worth mentioning that we also experimented with adding more heuristics in Phase~1, e.g., mapping multiple product names of a brand to a single brand name (Outlook to Microsoft), mapping old brand names and aliases to the latest (Facebook to Meta), etc. Yet, the performance did not match the system in both phases. 

\subsection{Interpretation of phishing classification (\textbf{Q5})}
\label{subsec:interpretation}

We now demonstrate the explanations provided by LLMs for phishing detection in both phases. 

\paragraph{\textbf{Brand identification by first-phase LLM}} 
Since ML-based solutions are prone to false positives, there is a need to provide an interpretation that might better help the user (e.g., a browser warning with an explanation). 
Therefore, in our design, the first-phase LLM gives the identified brand and supporting evidence as output, which are passed on to the second-phase LLM. The second-phase LLM is instructed to provide the final results along with interpretable evidence. The following example shows the brand identification result from GPT-4 against a real phishing webpage targeting WhatsApp. In this case, we provide both HTML and screenshots as input to the model for analysis. We observe that the LLM considers not only logo images and texts in the header and title, but also the brand color for its decision.

\vspace{0.2cm}
\begin{cooltextbox2} 
\small{
\texttt{\textbf{Brand}: WhatsApp}\\
\texttt{\textbf{Has Credentials}: Yes}\\
\texttt{\textbf{Has Call-To-Actions}: Yes}\\
\texttt{\textbf{List of Credentials fields}: Phone number}\\
\textbf{List of Call-To-Actions}: INSTALL UPDATE\\
\textbf{Confidence Score}: 9.50\\
\textbf{Supporting Evidence}: The screenshot prominently displays the WhatsApp logo along with a characteristic green color scheme, which is strongly associated with the brand.
The title in the HTML information, ``WhatsApp,'' the meta description, and headers text also merely state ``WhatsApp,'' reaffirming the identity of the webpage as belonging to WhatsApp. \dots }
\end{cooltextbox2}

\paragraph{\textbf{Phishing classification by second-phase LLM}} The second-phase LLM is introduced to resolve the gap between the brand identified (by the first-phase LLM), the domain name, and the actual brand targeted by the attack.  The following is an example of the input and output of the second-phase LLM for domain verification. We recall that our instruction specifically asks the LLM not to visit the URL so that our system is not infected via the malicious page (through various means, including new prompt injection attacks that target LLMs as discussed in Section~\ref{sec:discussion}). 

\vspace{0.2cm}
\begin{cooltextbox2} \small {
\textbf{Input}\\
\textbf{- URL}: https://polert.xyz/52Lp/whatsapp.html\\
\textbf{- Identified Brand}: WhatsApp\newline

\noindent \textbf{Output}\\
\textbf{- Genuine/Phishing}: Phishing\\
\textbf{- Evidence}: The URL uses the domain ``polert.xyz'' instead of the official domain ``whatsapp.com'', which suggests it is not associated with the official WhatsApp brand and may be suspicious or malicious.\\
\textbf{- Genuine URL}: https://www.whatsapp.com}
\end{cooltextbox2} 

We observe that the second-phase LLM compares the domain name to the identified brand (given at input), while also suggesting what might be the genuine webpage URL. In other examples, we noticed that the second-phase LLM is also able to correct misleading identification from the first-phase LLM, such as typos due to failed OCR. For example, the webpage of `Credit Agricole' was wrongly identified by Gemini as `\texttt{Credit Agricoole}' but the second-phase LLM detected this typo as `\texttt{Credit Agricole}'. 
The strict and specific instruction for the second-phase LLM helps to keep the focus on brand and domain matching and correct the LLMs' non-deterministic nature.

\subsection{Exclusive value of each LLM (revisiting \textbf{Q2} and \textbf{Q3})}
\label{subsec:exclusive-LLM}

\begin{figure}
    \centering    
    \includegraphics[width=0.47\textwidth]{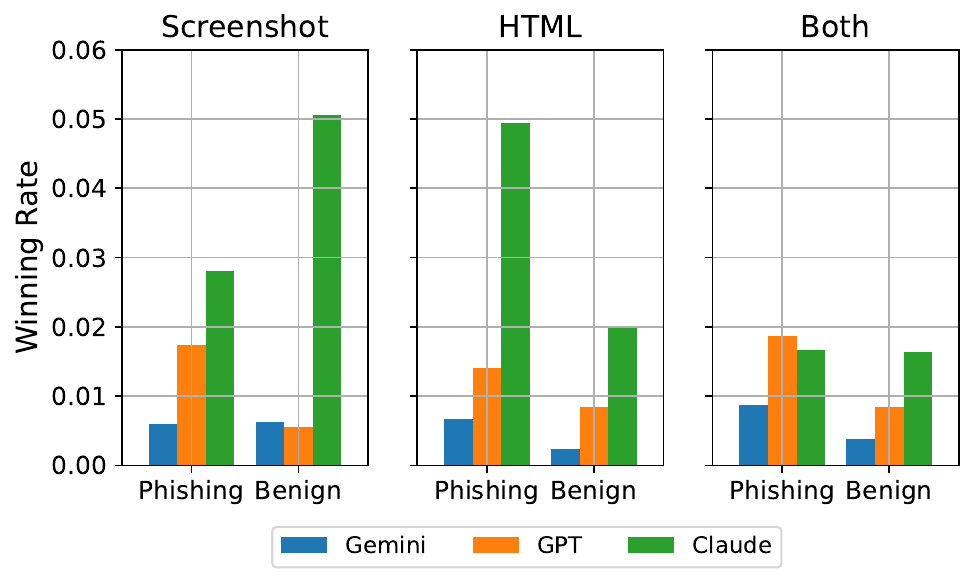}
    \caption{Exclusive true (winning) cases between LLMs}\label{fig:exclusive_win}
\end{figure}

Earlier, we observed that GPT-4 performed better than the other two LLMs; we now go deeper into the analysis. In particular, we are interested in analyzing the {\em exclusive} winning rate of each LLM in detecting phishing pages.
{Figure~\ref{fig:exclusive_win}} plots the percentage of phishing pages correctly identified by only one LLM, which other models failed to detect.

As shown in the leftmost plot for webpage screenshots in {Figure~\ref{fig:exclusive_win}}, Claude is able to exclusively detect more than 5\% of the benign web pages and close to 3\% of phishing pages.  The plot in the middle for HTML input shows that nearly 5\% of phishing webpages that Gemini and GPT could not detect are successfully detected by Claude. This exclusive identification capability reduces to less than 2\% when two types of input data are given (see the rightmost plot in {Figure~\ref{fig:exclusive_win}}).  This result shows the value of combining both the input types, which might not be obvious from the precision and recall metrics (Figure~\ref{fig:perform_llm_fig}). Another likely interpretation is that, if only a webpage screenshot is used for phishing detection, and the budget allows the use of two LLMs, GPT and Claude together, it forms the better option based on this empirical study. 

\begin{table*}[ht]
    \caption{Data effect category for identification results}
    \centering    
    \begin{tabular}{@{\hspace{1em}} l @{\hspace{1em}} | @{\hspace{1em}}  c @{\hspace{1em}} c @{\hspace{1em}} | @{\hspace{1em}} c @{\hspace{1em}} c @{\hspace{1em}} | @{\hspace{1em}} c @{\hspace{1em}} c @{\hspace{1em}}}
    \toprule
     Input & \multicolumn{6}{c}{Brand identification result} \\
    \midrule
    SS    & False & True  &  True & False & True &False  \\
    HTML  & True & False &  False & True & True & False \\
    Both  & False & False &  True & True & False & True \\
    \midrule 
    Category & \textbf{\texttt{Negative SS}} & \textbf{\texttt{Negative HTML}}  &
    \textbf{\texttt{Relying on SS}} & \textbf{\texttt{Relying on HTML}} & \textbf{\texttt{Conflict}} & \textbf{\texttt{Synergy}}\\
    \bottomrule
    \end{tabular}
    \label{tab:condition_input_copare}
\end{table*}

\subsection{Conflict and Synergy from input data (\textbf{Q6})}~\label{sec:eval:conflict}

\begin{figure}
    \centering
    \subfloat[Single input reliance]{\label{fig:input_reliance}
        \includegraphics[width=0.23\textwidth]{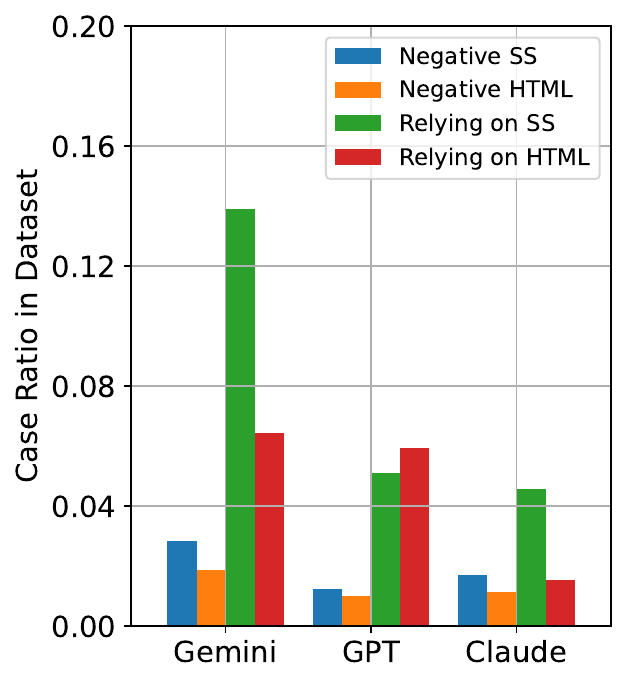}}
    \subfloat[Conflict and synergy]{\label{fig:input_crosseffect}
        \includegraphics[width=0.23\textwidth]{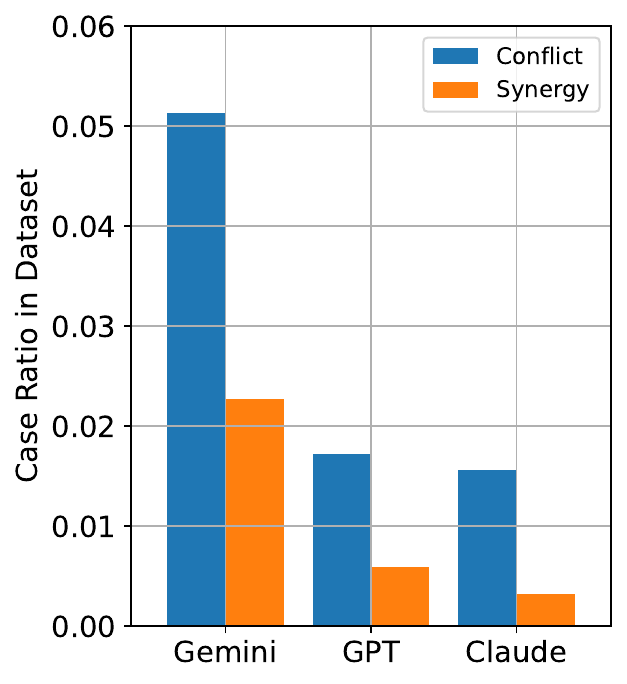}}
    \caption{Input data reliance and cross-effects}\label{fig:conflict_synergy}
\end{figure}

Another interesting question is whether one input type influences the identification result negatively, not just positively~(\textbf{Q6}). Table~\ref{tab:condition_input_copare} defines the corner cases of input combinations; e.g., `Negative SS' category means that a screenshot negatively affects the performance and providing only HTML is better.
This helps us to analyze some exceptional cases in which an LLM returns the correct identification only when limited information is given and, in particular, gives a wrong answer when provided with another input type.
Figure~\ref{fig:input_reliance} presents the ratio of phishing pages for these corner cases. 

The left-most column, \texttt{Negative SS}, in {Figure~\ref{fig:input_reliance}} shows that the brands of around 2.8\% of samples in our dataset are correctly identified by Gemini when we give {\em only} HTML  input, but gets the brand wrong when screenshot is provided as additional input. That is, neither screenshot alone nor screenshot and HTML together help in identifying these samples.  In contrast, for nearly $14\%$ of samples, Gemini relies on the screenshots (\texttt{Relying on SS}), which means it returns a correct brand when a screenshot is given (with or without HTML),  but the wrong brand when given only HTML information.

Interestingly, the composition of multiple inputs not only creates synergy but also creates conflict (defined in Table~\ref{tab:condition_input_copare}).   Figure~\ref{fig:conflict_synergy}(b) illustrates this. 
With Gemini, nearly 5\% of the samples fall under the  \texttt{conflict} category, resulting in the wrong answer, although the correct brand is identified when provided with either of the input type {\em independently}.  On the other hand, around 2\% of samples found correct answers in the \texttt{synergy} category, that otherwise failed with one input type (screenshot or HTML). The \texttt{conflict} samples are less for GPT and Claude, with the latter not even finding the synergy contributing significantly to the detection performance.

\subsection{Comparison with state-of-the-art brand-based phishing detection ({\textbf{Q7}})}
\label{sec:eval:vpn}

\begin{table}[t]
    \caption{System performance comparison dataset}
    \centering
    \begin{tabular}{@{\hspace{1em}} l @{\hspace{1em}} | @{\hspace{1em}}  c @{\hspace{1em}} c @{\hspace{1em}} }    
    \toprule
      & Training (brands) &  Testing (brands)  \\
    \midrule
    Benign         & 1270 (34)  & 835 (829)  \\
    Phishing       & 315 (34)   & 612  (34)   \\    
    \midrule
    Total          & 1585 (34)  & 1447 (863)  \\
    \bottomrule
    \end{tabular}
    \label{tab:system_dataset}    
    \vspace{-0.5cm}
\end{table}

In this section, we compare the LLM-based phishing detection with a state-of-the-art vision-based phishing detection system, i.e., VisualPhishNet~\cite{abdelnabi2020visualphishnet}.

In VisualPhishNet, the model is trained using the screenshots of the webpages of a list of protected brands. The assumption is that the (small) list of protected brands is the ones that are often targeted by attackers and need to be protected against. Therefore, the model aims to discriminate between the phishing pages targeting the protected brands and the other benign pages for the unprotected brands (a white list easily detects the legitimate pages corresponding to the protected brands).  The phishing webpages targeting brands other than the protected brands are out of the scope of VisualPhishNet.

We select a subset of our data for the evaluations here.
Importantly, we select the target brands that are there in both the benign and phishing classes of our dataset~{Section~\ref{sec:data}} so that VisualPhishNet sees all target brands in its training. This filtering reduces the benign and phishing pages to 1585 and 927, respectively. We split these into training and test such that all brands are included in the training phase, see {Table~\ref{tab:system_dataset}}.
To be sure, in the testing phase, we evaluate the detection performance against the phishing pages that target the protected brands and the benign pages not included in the protected brands. Since target brands of phishing pages fall in the protected list, the brand identification directly leads to the detection of phishing pages. 

\begin{figure}
    \centering
    \subfloat[Precision]{\label{fig:vpn_prec_fig}
        \includegraphics[width=0.23\textwidth]{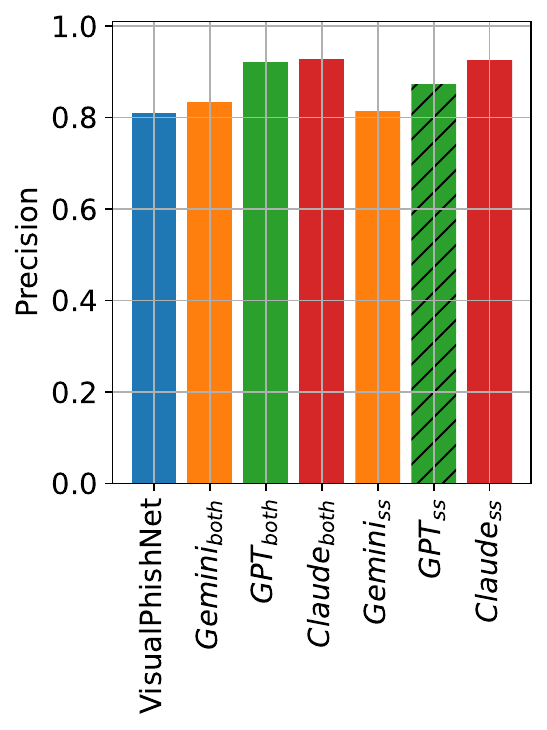}}
    \subfloat[Recall]{\label{fig:vpn_recall_fig}
        \includegraphics[width=0.23\textwidth]{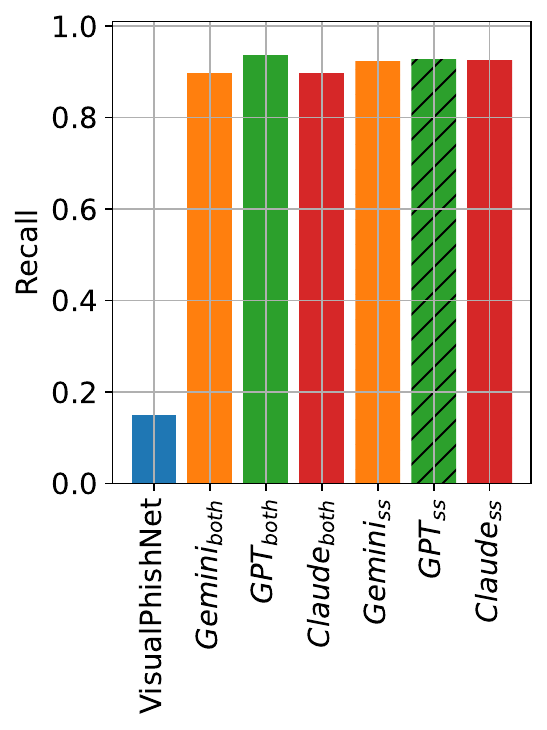}}
    \caption{Precision and Recall of brand identification}\label{fig:perform_vpn_fig}
\end{figure}

{Figure~\ref{fig:perform_vpn_fig}} shows the brand identification performance between VisualPhishNet and the multimodal LLMs.  Note that the performance of LLMs differs from the ones in {Figure~\ref{fig:perform_llm_fig}} due to the difference in the datasets.  One clear observation is the low recall of VisualPhishNet in {Figure~\ref{fig:vpn_recall_fig}}.  When we set a detection threshold for VisualPhishNet so that it achieves a similar and comparable level of precision as the LLMs, its recall is less than 15\%. Whereas the LLMs achieve a recall (detection rate) of around 90\%.

VisualPhishNet's low detection accuracy (recall) indicates its limited efficacy in identifying phishing webpages in challenging scenarios. 
High deviation of phishing webpages from their target benign counterparts leads to incorrect brand identification, resulting in low recall.
In contrast, multimodal LLMs demonstrate much better performance. 
These models possess multiple advantages for brand identification, such as the ability to extract and analyze text without relying on visual similarity to benign webpages. This capability extends detection coverage to a wide range of brands, including those beyond the protected set.

\subsection{Cost and Overhead ({\textbf{Q8}})}~\label{sec:eval:cost}
In this section, we analyze the cost of utilizing multimodal LLMs for the brand identification task.  As of now, the most powerful and capable LLMs are the commercial service models backed by substantial resource investment.  The effectiveness of these models naturally comes with a cost based on the size of input and output data, usually measured in the number of \textit{tokens}.  The price per token varies among service providers such as OpenAI, Google, and Anthropic, and depends on user contracts. However, the distribution of token consumption is dependent on the specific task. These elements influence both the feasibility of using LLMs for brand identification and their susceptibility to economic attacks.

We consider the brand identification experiments for the dataset in Table~\ref{tab:lmm_dataset} with the prompts defined in {Section~\ref{sec:system}}. {Figure~\ref{fig:token_consumption}} illustrates the total token consumption---tokens for input and output of a request---for this set of experiments.
By comparing the three box-plots in {Figure~\ref{fig:token_consumption}} for the different input data (screenshot, HTML, and both), we can generally observe an expected trend that larger input requires more tokens. However, we also observe different token consumption characteristics of the models.  Although all three models are tested with identical input and prompt, the distributions of the tokens vary.  

\subsubsection{Token consumption for screenshot images}
\begin{figure}
    \centering    
    \subfloat[Screenshot]{\label{fig:token_ss}
        \includegraphics[width=0.16\textwidth]{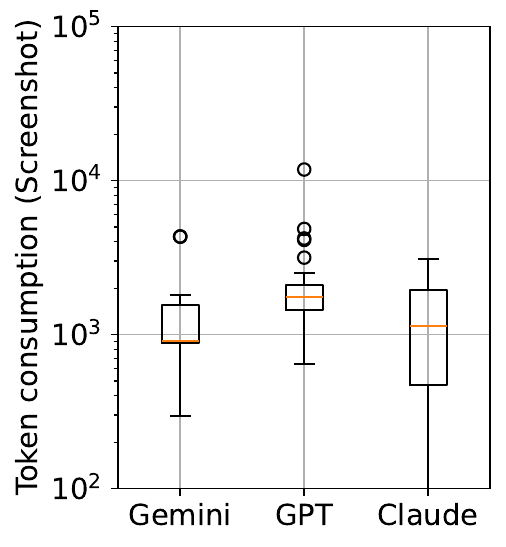}}
    \subfloat[HTML]{\label{fig:token_html}
        \includegraphics[width=0.16\textwidth]{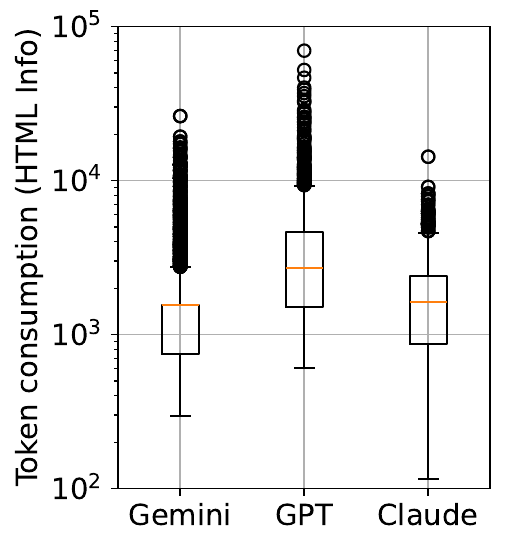}}
    \subfloat[Both]{\label{fig:token_both}
        \includegraphics[width=0.16\textwidth]{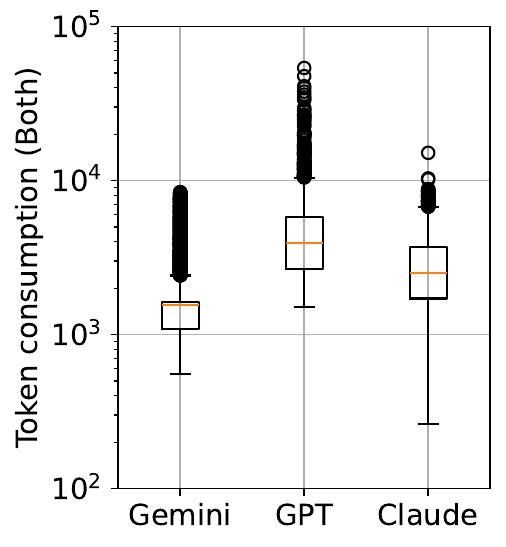}}        
    \caption{Token consumption based on the input data }\label{fig:token_consumption}
\end{figure}

{Figure~\ref{fig:token_ss}} shows that, for screenshots, the tokens consumed by Claude3 are more widely distributed in comparison to Gemini and GPT. Note that the Y-axis is in the log-scale. 
Whereas the top 25\% expensive samples (around $2.3\times10^3$) are less than twice of the low 25\%  samples (around $1.7\times10^3$) in GPT-4, 75\% high samples in Claude3 (around $2\times10^3$) is four times expensive than low 25\% (around $5\times10^2$) sample.  Though we cannot directly compare the price of one token between the models, the result clearly shows that Claude3 model has a wider range of cost risk with unpredictable token consumption.

\subsubsection{Token consumption for text input}
In contrast to image input, token consumption for HTML, which is text-based, shows more consistency across models; see {Figure~\ref{fig:token_html}(b)}.  However, we observe far more outliers than in the image-only cases because the input length for text input (and therefore the corresponding tokens) is much more flexible than for images; images have a size restriction and an auto-resizing policy.  While the highest outliers in screenshot input are five to six times the mean, the top outlier in the HTML input is as many as thirty times the mean.

\subsubsection{Token consumption for both image and text input}
Input with the image and text (HTML) together results in higher token consumption than any one single type. The trend (Figure~\ref{fig:token_both}) extends from the observations we made with single input types. 

When considering phishing webpage detection systems that accept public inputs, the uncertainty of cost (of using LLMs) against arbitrary inputs is a critical factor to consider. This risk should be managed to avoid unwanted costs. The three models we evaluated do not provide any built-in input token limitations but only output limitations. The token consumption for images (webpage screenshots) can be computed based on the predefined token calculation formula according to image size. However, for text input, the models calculate actual token consumption only after processing the prompts. As discussed in {Section~\ref{sec:discussion}}, this has implications for adversarial attacks.

\begin{cooltextbox}
\textsc{\textbf{Takeaways.}}
\\
i)~LLM-based phishing detection achieves a high detection rate at high precision. GPT and Claude perform better than Gemini in our experiments. \\ 
ii)~While providing both the screenshot and HTMP contents of a webpage to an LLM gives the best results, with screenshot, we have the advantage of controlling cost in terms of tokens consumed. When a webpage screenshot is the only input, Claude serves as the best of the three LLMs we evaluated for phishing detection. \\ 
iii)~LLM-based phishing detection system performs much better than VisualPhishNet, a state-of-the-art brand-based phishing detection solution, demonstrating the capability of multimodal LLMs to identify well-known brands via multiple invariants such as logos, theme, favicon, etc.   
\end{cooltextbox}

\section{Case Study}
\label{sec:case-study}

\subsection{False Identification}
\label{sec:falseidenfify}
In this section, we review the false brand identification cases by multimodal LLMs.  According to our manual investigation, most of the false identifications (of first-phase LLM) could be corrected by the second-phase LLM.  However, these false identifications also point out the weakness of LLMs, which attackers can intentionally exploit (refer Section~\ref{sec:discussion}). 

\begin{figure}
    \centering
    \frame{
    \includegraphics[width=0.35\textwidth]{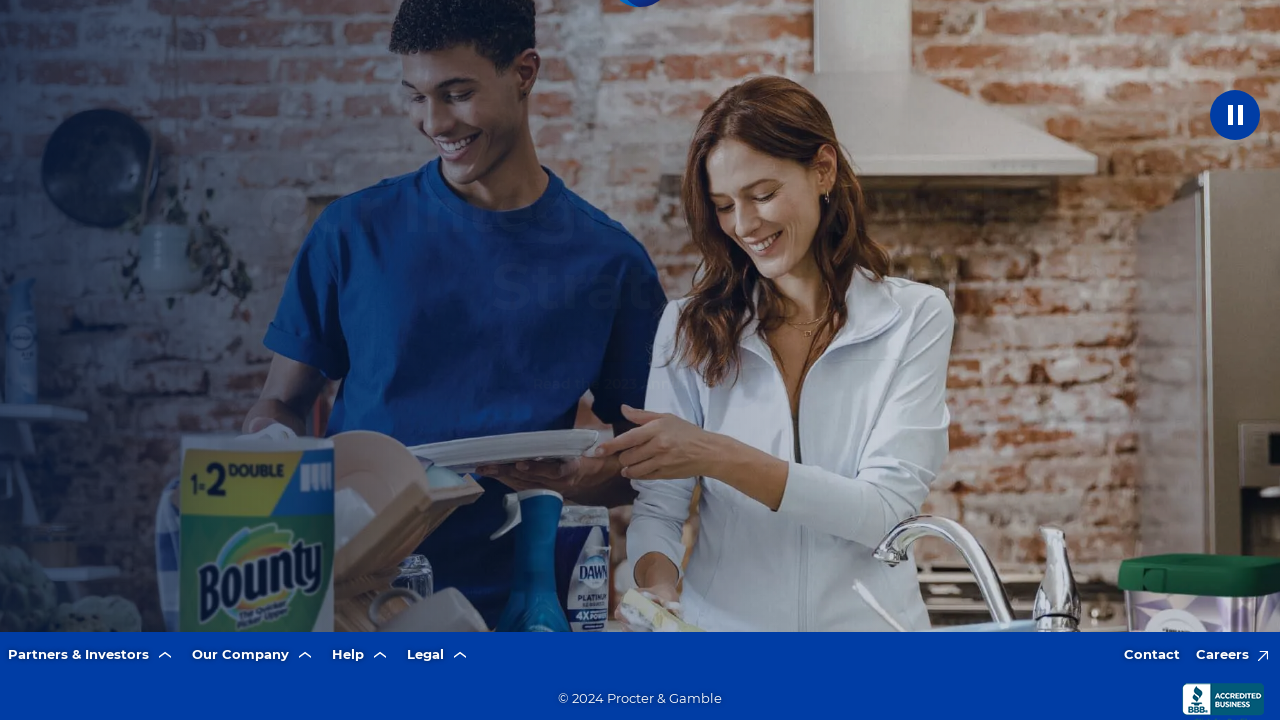}
    }
    \caption{Misidentification case: detected as \textit{Bounty}}\label{fig:ss_fp_wrong_capture}
\end{figure}

\begin{figure}
    \centering    
    \frame{
    \includegraphics[width=0.35\textwidth]{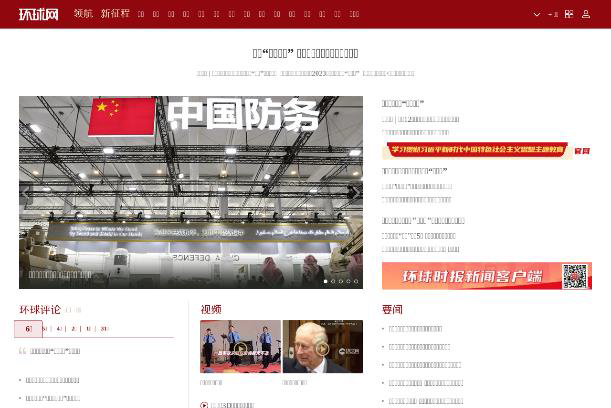}
    }
    \caption{Misidentification case: Brand in a local language}\label{fig:ss_fp_non_english}
\end{figure}

\begin{itemize}

\item {\bf Misreading information on screenshots}: Wrong logo detection may happen when there is another (potentially) prominent logo on the screenshot, especially when no HTML data is given to the LLM.  {Figure~\ref{fig:ss_fp_wrong_capture}} is an example illustrating this, wherein the LLM returns `\textit{Bounty}' for the webpage of \textit{P\&G (Procter \& Gamble)}. These cases are resolved by providing HTML data along with the screenshot.\\

\item \textbf{Misreading information on texts}: 
In contrast to the above scenario, there are also cases where the correct brand is present on the screenshot, but a wrong name in the text is identified as the brand by an LLM.  The texts in the HTML contents of a webpage contain much more information than in the screenshot and, in particular, might mislead the LLM to wrongly identify the brand.  Unlike screenshots, a text has more chance to intentionally highlight the importance of a specific keyword by coming up with strong contexts, e.g., `company,' `brand,' `official,' `contact,' etc. And their repetition also appears to mislead an LLM.  This could be exploited for indirect prompt injection attack (refer Section~\ref{sec:discussion}).

\item  \textbf{Brand name in regional language}: Even when an  LLM correctly identifies the brand from screenshots or HTML data, in some cases, it returns the name in the language of the region where the webpage is mainly serviced.  Figure~\ref{fig:ss_fp_non_english} shows an example with the brands in a regional language.  Even though LLMs have language detection and translation capability, there are multiple challenges here, e.g., when a regionally relevant word is written in English, when non-English words/brands are used, etc. However, technically, the output of the first-phase LLM is correct; therefore, what is necessary is to explore the possibility of giving more context to the second-phase LLM to do better matching. 

\end{itemize}

\begin{figure}
    \centering    
    \frame{
    \includegraphics[width=0.4\textwidth]{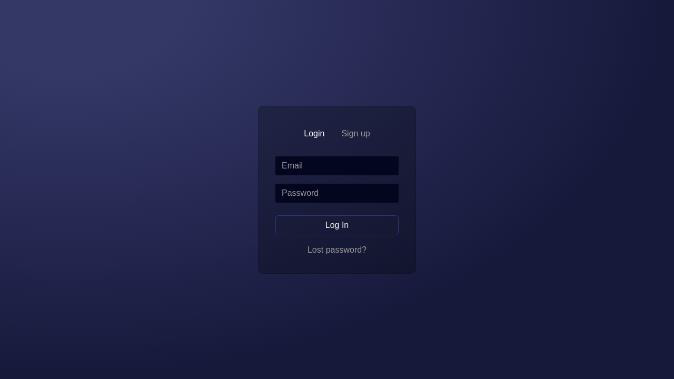}
    }
    \caption{Misidentification case: no clear indicator of brand}\label{fig:ss_fp_no_indicator}
\end{figure}

There might also be cases where the phishing pages have no information related to any brand, Figure~\ref{fig:ss_fp_no_indicator} is an example. Yet, such a phishing page still might be effective if it is part of a multi-stage phishing attack, where the brand was used to deceive victims in an earlier stage (e.g., email, webpage, etc.). 

\subsection{Error Cases}\label{sec:eval:error}
LLMs often return an error response to the request along with their internal policies, e.g., risky contents, privacy, and input size limitation.
They also return an error when the given webpage is detected by its safety check filter as having harmful content, such as sexually explicit content, hate speech, harassment, or other dangerous content. It is not surprising that a phishing webpage may have harmful content. However, this is not a deterministic behavior, and if we try multiple visits, the output might vary depending on the LLM. 

In our experiments, we found Gemini to be stricter than GPT and Cluade when it comes to safety filters. This adversely affects the detection capability in the case of phishing.   

Lastly, all LLMs occasionally return an internal server error with a message. This error is not related to input data but rather due to the service stability, causing a delay in inference. 

\section{Robustness against Adversarial attacks}\label{sec:eval:adv}
We conduct experiments to answer the following  question: \\

\noindent \textbf{Q9.} Is an LLM-based phishing detection system robust against adversarial attacks?

\subsection{Adversarial attacks against visual-based phishing detection}
The adversarial attack is one of the known effective attack vectors to deceive brand-based phishing detection systems using CV (Computer Vision) models. In~\cite{lee2023attacking}, it is shown that the perturbed images (logos) with noise effectively hinder the brand identification capabilities of CV models while also deceiving users~\cite{lee2023attacking}.  More recently, Yuan {\it et al.} report that most of the perturbation techniques in images do not increase or decrease the user's phishing recognition, yet half of the users failed to detect the phishing pages~\cite{yuan2024are}.  We examine the effectiveness of the adversarial attack against LLMs with the same image datasets that are used by these studies~\cite{lee2023attacking,yuan2024are}.

\subsubsection{Adversarial Attack~1 --- Webpage source perturbation}
Yuan {\it et al.} introduce four webpage source perturbation techniques in their study~\cite{yuan2024are} and show their effects on human users. According to their user study, changes in webpage components, such as images and texts, which have been shown to evade ML-based phishing webpage detection systems in various studies, still result in a phishing success rate of over 40\% among users. Specifically, they used four perturbation techniques: altering the background image, modifying the footer image, introducing typos in brand name text, and unveiling the password form.

We test the 90 perturbed phishing webpage samples from~\cite{yuan2024are}, targeting 15 popular brands and 28 wild-perturbed samples (real perturbed phishing samples from the wild); these were evaluated in their user study as mentioned above. In addition to the dataset in Yuan {\it et al.} study~\cite{yuan2024are}, we add the genuine webpages of the 15 target brands to the test set.  In total, we test 133 samples---105 for 15 brands, and 28 wild-perturbed phishing webpage samples.

More specifically, the dataset has two unperturbed phishing webpage samples for each brand.  The in-lab perturbation sample set is created by adopting four perturbation techniques for the phishing samples.  Two perturbation techniques (out of the four) are applied to one sample of a brand, and the other two techniques are applied to another sample of the brand so that each brand has four perturbed samples. Lastly, twenty-eight perturbed phishing samples in the wild are used to compare the in-lab and the wild perturbed samples.

\subsubsection{Adversarial Attack~2 --- Noise perturbation on Logo}
Based on the observation that most recent visual-based brand identification solutions rely heavily on the logo image of a webpage to detect its brand, Lee~{\it et~al.} focused on perturbing the logos of popular webpages using generative models. According to their empirical study, the perturbed logo images successfully evade the logo-identification capability of the state-of-the-art phishing detection models, while they are not effectively recognized by human users~\cite{lee2023attacking}.
For instance, noise-perturbed logo images generated by the ViT model~\cite{dosovitskiy2020image} could drop the detection rate of the brand identification models from over 90\% to less than 60\% and, in the worst case, to nearly 5\%, at the same false-positive rate.  Furthermore, the noise-perturbed logo images are not recognized by humans in their user study with nearly 300 participants.

To examine the effect of the noise-perturbed logs against the brand identification capability of LLMs, we test a selected subset of the dataset used in~\cite{lee2023attacking}.  We select 10 perturbed logo samples per brand for the same 15 brands used in the (above) webpage source perturbation test.

\subsection{Results}
\begin{table}[t]
    \caption{Brand identification against adversarial samples}
    \centering    
    \begin{tabular}{l|c c c| c c c }    
    \toprule
                & \multicolumn{3}{c}{Webpage source~\cite{yuan2024are}} & \multicolumn{3}{c}{Noise on Logo~\cite{lee2023attacking}} \\
                & Correct & Wrong & Error & Correct & Wrong & Error \\
    \midrule
    Gemini   & 130 & 1  & 2 & 138& 5 & 7 \\
    GPT      & 133 & 0 & 0 & 146 & 3 & 1 \\
    Claude   & 128 & 0 & 5 & 147 & 1 & 2 \\
    \bottomrule
    \end{tabular}
    \label{tab:adversarial_result}
\end{table}

{\bf Table~\ref{tab:adversarial_result}} presents the results of brand identification experiments against the adversarial samples.  While it is clearly showing that the robustness of LLMs against adversarial attacks.

\paragraph{\textbf{Adversarial Attack~1 --- Webpage source perturbation}}
In the brand identification test with 133 images that include screenshots of the original, phishing, lab-perturbed, and wild-perturbed 15 popular webpages~\cite{yuan2024are}, the brands of all 133 screenshots are correctly identified by GPT and Claude demonstrating its capability against webpage source perturbation. GeminiPro made a single failure due to the presence of multiple logos, which, however, was not forced by the attack.
In the experiment, we also observed a case where the second-phase LLM helped.
This sample had eBay with a (forced) typo and was wrongly identified in the first phase as \textit{WBay} (a television channel in the United States). However, the second-phase LLM could correct the brand to eBay. 

\paragraph{\textbf{Adversarial Attack~2 --- Noise perturbation on Logo}} The results in {\bf Table~\ref{tab:adversarial_result}} show that noise-based adversarial attacks are more effective against the LLMs than perturbation on web sources, although the difference is not very significant. We observe only five or fewer false identification cases from all the three LLMs.

Note that, the logo perturbations in~\cite{lee2023attacking} are created automatically using generative adversarial perturbations~\cite{poursaeed2018generative} trained to evade CV models used for logo identification (Siamese, ViT, and Swin). Thus, in contrast to the manually crafted webpage perturbation (in the first attack), the noise perturbation of logos (in the second attack) is automated and targeted. This might explain the relatively better evasion capability of the second attack (noise perturbation of logos). 

\paragraph{\textbf{Strength of LLM against Adversarial attacks}}
The adversarial phishing attacks described above~\cite{lee2023attacking, yuan2024are} target previous research works that used CV models (e.g., Siamese model) for detecting the similarity of a given webpage with webpages~\cite{abdelnabi2020visualphishnet} and webpage logos~\cite{lin2021phishpedia, liu2022inferring} of well-known legitimate brands; and these models are shown to be weak against the attacks. 
However, in our experiments, where we evaluate the same two attacks against LLM-based phishing detection, we observe that the LLMs are robust against these adversarial attacks. We postulate the following reasons for this difference in performance: (i)~the CV models in~\cite{abdelnabi2020visualphishnet, lin2021phishpedia, liu2022inferring, lee2023attacking, yuan2024are} are trained using collected datasets consisting of only a few thousand of pages. Therefore, deviations in webpages or logos might evade these models trained on limited data. ii)~On the other hand, multimodal LLMs are pretrained with massive data of various types (text, images, videos, etc.) and are therefore more robust to perturbations. Furthermore, while CV models focus on high similarity within specific invariants (webpage/logo) they are trained on, pretrained LLMs are capable of analyzing multiple aspects of a webpage (e.g., logos, input form, theme, favicon, and so on).

\paragraph{\textbf{Weakness of LLM against Adversarial attacks}}
Based on the failed samples, the following seems to be the weakness of LLMs.  First, effectively applied noise perturbations hinder the OCR of LLM. This was particularly evident when the text on the image was not in English. Another potential vulnerability is when an LLM lacks sufficient background knowledge of the genuine brand, meaning the brand is not globally popular. It remains to be tested if LLMs can identify regionally relevant brands, which is also a common weakness observed in brand-based phishing detection systems~\cite{li2024knowphish}.

\section{Discussion on new attack vectors due to the use of LLMs}\label{sec:discussion}

\textbf{Q10.} When LLM drives the capabilities of a phishing detection system, what are the new attack vectors that arise? We discuss them below. 

\paragraph{\textbf{Threat due to open and accessible model}}
The LLMs we use for brand identification are openly available to everyone, including attackers. This accessibility allows attackers to generate and test adversarial examples (page contents and their screenshots) against these open-access LLMs.The successful samples could lead to the development of phishing pages that successfully evade LLM-based phishing systems.

\paragraph{\textbf{Economic denial-of-service attack}}
Attacks on a detection system not only aim at inaccurate detection. An adversary may also attempt to increase the overhead and operational cost of an LLM-connected system. Even if a system has control over the LLM token consumption and rate limiting, a series of crafted inputs hitting the upper bound of the rate-limiting will cause persistent economic damage to the system operator. See Section~\ref{sec:eval:cost} for an analysis of the cost of LLMs in our study.

\paragraph{\textbf{Indirect prompt injection}}
LLMs are vulnerable to prompt injection attacks that result in harmful or unintended output. Recent research shows that images can be perturbed so as to adversarial instructions that are not visible to users, but causes the LLMs to generate harmful output, e.g., a link a malicious page~\cite{indirect-prompt-injection-Vitaly-2023}. Attackers may deceive ways to carefully craft instructions in phishing pages that enable evasions. For example, extending the ideas for attacks proposed in~\cite{wu2024wipi}, an attacker could emphasize the domain name of the phishing URL within the website in an obscure way that fools the LLM, while making it imperceptible to users. Defending against this may require guardrails or keeping the instruction channel separate from the data channel. 

\paragraph{\textbf{Obsolete or missing knowledge}}
A known weakness of LLM is outdated information, as the model may not have access to the correct information about the brands in real-time. Therefore, based on the training cut-off time, an LLM might not be aware of the latest brands. Indeed, it also remains to be tested if LLMs are knowledgeable of regional brands that might not be globally well-known but yet target (victims) in a particular region.

\paragraph{\textbf{Training data poisoning}} LLMs are typically trained on data collected from a wide range of sources across the internet. An attacker could systematically insert harmful or misleading content into these sources, hoping that the corrupted data will be included in the training set.  

For example, an attacker might insert biased content into widely-read blogs, forums, or other high-visibility platforms.  Over time, these poisoned examples could lead the model to produce outputs that reflect those biases or errors.  
\section{Conclusions}
We proposed and comprehensively studied an LLM-based two-phase phishing detection system. Our evaluations showed the efficacy of multimodal LLMs in detecting phishing pages while being robust to known adversarial attacks. 
However, LLMs themselves give rise to new attack vectors. As a next step, we plan to study how an LLM-assisted phishing detection system can be hardened against these new attacks.

\bibliographystyle{IEEEtran}
\bibliography{eCrime-main}

\end{document}